\documentclass[%
 reprint,
 amsmath,amssymb,
 aps,superscriptaddress,
]{revtex4-2}

\usepackage{xcolor}
\usepackage{graphicx}
\usepackage{dcolumn}
\usepackage{bm}

\usepackage{caption}
\usepackage{subcaption}

\begin{document}

\preprint{APS/123-QED}

\title{Witnessing quantum steering by means of the Fisher information}

\author{Ilaria Gianani}
\affiliation{Dipartimento di Scienze, Universit\`a degli Studi Roma Tre, Via della Vasca Navale 84, 00146 Rome, Italy}

\author{Vincenzo Berardi}
\affiliation{Dipartimento Interateneo di Fisica ``Michelangelo Merlin'', Politecnico di Bari, Via Orabona 4, 70126 Bari, Italy}

\author{Marco Barbieri}
\affiliation{Dipartimento di Scienze, Universit\'a degli Studi Roma Tre, Via della Vasca Navale 84, 00146 Rome, Italy}
\affiliation{Istituto Nazionale di Ottica, CNR, Largo Enrico Fermi 6, 50125 Florence, Italy}


\begin{abstract}
Capturing specific kinds of quantum correlation is of paramount importance for quantum networking. Different routes can be taken to achieve this task, highlighting different novel aspects of such quantum correlations. Following the recent theoretical results by Yadin, Fadel and Gessner [Nat. Commun. 12, 2410 (2021)], we demonstrate experimentally how steering manifests in the metrological abilities of a bipartite state. Our results confirm the relevance of this novel approach, and compare the outcome with already employed alternatives. 

\end{abstract}

\maketitle

\section{Introduction}
Understanding the potential of processing information by means of quantum systems is bringing about an intense technological effort. This goes under the name of second quantum revolution~\cite{doi:10.1098/rsta.2003.1227}, to tell it apart form the first one that developed the new theory in the early twentieth century. The current technology-oriented approach bears conceptual implications as well: under the new light, quantum properties are considered not only for their physical meaning, but also for their implications in information processing tasks.  

This technology-oriented mindset has allowed to recognise the existence of different categories of quantum correlations, distinguished according to their operational meaning~\cite{app9245406}. When observing a bipartite system, one may ask different questions: whether measurement outcomes could be reproduced by means of local realistic models, leading to the concept of quantum nonlocality~\cite{RevModPhys.86.419};  whether correlations can be explained only in terms of local quantum states (steering)~\cite{RevModPhys.92.015001}; whether there exist a procedure to generate the state locally (entanglement)~\cite{RevModPhys.81.865}. These correlations, in turn, provide means for quantum communications at different levels of security and trust among the nodes~\cite{PhysRevLett.97.120405,PhysRevLett.98.230501,Masanes:2011th}, within the paradigm adopted for this categorisation~\cite{PhysRevLett.108.200401}.

Steering captures in rigorous terms the effect underlying the Einstein-Podolsky-Rosen (EPR) paradox~\cite{PhysRev.47.777}, as originally highlighted by Schr\"odinger~\cite{schrodinger_1935}, and then formalised by Wiseman et al.~\cite{PhysRevLett.98.140402}. Two partners, Alice and Bob, share a bipartite state, however Bob doubts that Alice can steer his local state better than what she could by local hidden states. There exist criteria for assessing the unsuitability of such models~\cite{PhysRevA.80.032112,PhysRevLett.114.060404,PhysRevA.81.022101,PhysRevA.87.062103}, many of which have been tested in experiments~\cite{Saunders:2010xw,PhysRevX.2.031003,Handchen:2012jl,Smith:2012zt,Wittmann_2012,Cavalcanti:2015qy,Kocsis:2015ee,PhysRevLett.116.160404,Westone1701230,PhysRevLett.121.100401,PhysRevLett.125.020404}. In particular, Reid's criterion states that steering can be revealed by the inappropriateness of Heisenberg's relations for the variances of Bob's observables, conditioned on his communication with Alice~\cite{PhysRevA.40.913}. Based on this approach, Yadin, Fadel and Gessner (YFG) \cite{YBG} have established a connection between the presence of steering, as captured by Reid's criterion, and the metrological power of Bob's conditional states. In this work, we present an experimental demonstration of use of the YFG criterion for the measurement of steering on two-photon states produced by non-classical interference. Our results show that albeit closely related to Reid's criterion, for our class of states the YFG approach delivers a  quantitative assessment of the amount of steering at the cost of increased resources being required. These results reinforce the employ of metrological figures beyond problems strictly related to sensing. 

\begin{figure}[b!]
\includegraphics[width=\columnwidth]{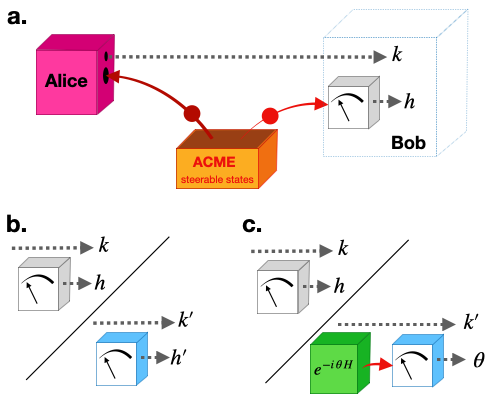}
    \caption{{\it Conceptual scheme.} a) Alice and Bob share a bipartite state. Alice claims she can steer Bob's state by performing a measurement of K with outcome $k$. Bob, receiving Alice's outcome and measuring his own, can check whether Alice is indeed steering his state or is pretending to do so by employing local hidden states. Bob can choose two different strategies: either  b) performing the required measurements for implementing Reid's criterion or  c) measuring the Fisher information of his states, conditioned by Alice's outcome.}
    \label{fig:concept}
\end{figure}

\section{Results} 
The concept of steering can be illustrated as follows (Fig.~\ref{fig:concept}). Alice and Bob have access to a bipartite state, and Alice claims that, by performing a measurement of $K=\sum_k k \vert k\rangle\langle k\vert$ and collecting the outcome $k$, she is able to steer the state of Bob's system, and guess the outcome $h$ of a measurement of his observable $H=\sum_h h \vert h\rangle\langle h\vert$. Bob, instead, is convinced this is not the case, and Alice merely mimics this process by means of a classical probability distribution for $k$, and the outcomes are determined by Born's rule applied to a local hidden state $\sigma^B_k(\lambda)$, whose probability $p(\lambda)$ is set by a classical variable $\lambda$. The joint statistics of $k$ and $h$ is described by means of the assemblage $\mathcal{A}(k,K)=p(k|K)\rho^B_{k|K}$, which links Alice's result $k$ for the measurement of $K$, and Bob's conditioned state $\rho^B_{k|K}$; the joint probability is then written as $p(k,h|K,H)=\langle h\vert\mathcal{A}(k,K)\vert h\rangle$. The local hidden state setting writes $\mathcal{A}(k,K)=\sum_\lambda p(\lambda) p(k|X,\lambda)\sigma^B_k(\lambda)$. Thus, the properties of the allowed assemblages are curtailed, specifically for what concerns the strength of their correlations, in analogy with what occurs with local hidden variables.

Steering is connected with the emergence of the EPR paradox~\cite{PhysRev.47.777}, as recognised in~\cite{PhysRevLett.98.140402}, in that the knowledge of Alice's value $k$ allows to make predictions on Bob's results beyond what would be allowed by Heisenberg's relation - a fact on which also Popper called to attention~\cite{Popper1985,Kim1999}. A way of quantifying these observation in a quantum state considers the average deviation of the actual value $h$ with respect to $\check{h}(k)$, the one predicted based on $k$:
\begin{equation}
\label{eq:deltaest}
    \Delta^2H_{\rm est} = \sum_{a,h} p(k,h\vert K,H)(\check{h}(k)-h)^2.
\end{equation}
When two incompatible measurements are carried out, a local uncertainty limit holds for non steerable states~\cite{PhysRevA.40.913,RevModPhys.81.1727}
\begin{equation}
\label{eq:reid}
    \Delta^2H_{\rm est} \Delta^2H'_{\rm est}\geq \frac{1}{4} \vert \langle [H,H'] \rangle \vert^2,
\end{equation}
where the average of the commutator is calculated on Bob's unconditioned reduced state $\rho^B$. Reid's criterion states that the violation of such an inequality may serve as a witness for an EPR paradox.

Inequalities in the form \eqref{eq:reid} are also found in quantum metrology when considering the estimation of the parameter $\theta$ in the transformation $e^{-i\theta H}$. The variance $\text{Var}(\theta)$ on the value of the parameter is limited by the Fisher information $F(\theta)$ associated to the quantum state, the chosen measurement, and the number $n$ of repetitions of the experiment according to the Cram\'er-Rao bound $\text{Var}(\theta)\geq1/(nF(\theta))$. The quantum Fisher information $F_Q(\theta)$, which is a function of the quantum state only, limits all possible Fisher information from above $F_Q(\theta)\geq F(\theta)$, which can be saturated for a particular measurement. The resulting inequality $\text{Var}(\theta)\geq1/(nF_Q(\theta))$ is known as the quantum Cram\'er-Rao bound~\cite{1976iv,Holevo:1414149,PhysRevLett.72.3439,Giovannetti:2011yq,PhysRevLett.96.010401,ParisReview,GLMreviewScience}. For pure states, $F_Q(\theta)=4\Delta^2H$, thus we can write the Cram\'er-Rao bound in a form reminiscent of Heisenberg's relation~\cite{PhysRevA.73.042307,PhysRevLett.98.160401}
\begin{equation}
\label{CRB}
    \text{Var}(\theta)\Delta^2H \geq \frac{1}{4n},
\end{equation}
which also applies for generic mixed states~\cite{ParisReview}.

\begin{figure}[h!]
\includegraphics[width=\columnwidth]{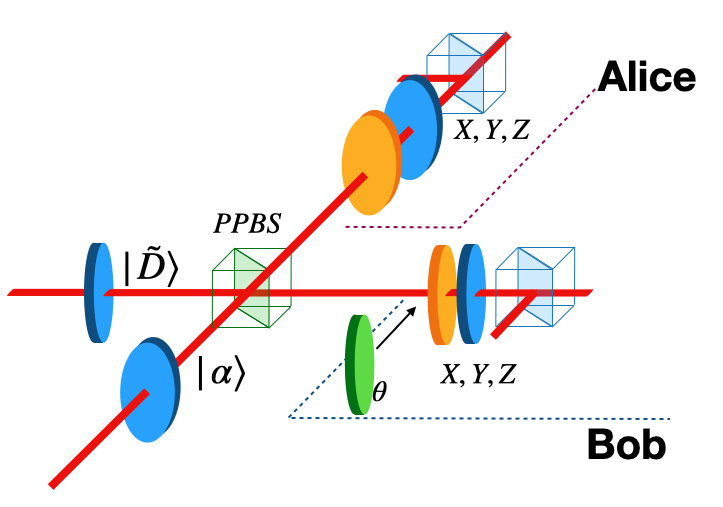}
    \caption{{\it Experimental setup.} Photon pairs are generated via spontaneous parametric downconversion with a CW 405 nm pump through a 3mm type I BBO crystal. The photons are then spectrally filtered (Full-width-half-maximum of 7.5 nm) and sent through single-mode fibres to the partially polarising beam splitter (PPBS): the input polarizations are controlled through two half-wave plates (blue), and take into account the necessary pre-bias to compensate for the different transmission probabilities of the partially polarizing beam splitter. Alice and Bob's measurements of the Pauli operators are then conventionally performed by means of a quarter wave plate (orange), half wave plate, and a polarizing beam splitter. For the measurement of the Fisher information, a further rotation is implemented on Bob's side by a half wave plate (green).}
    \label{fig:apparato}
\end{figure}
Combining results \eqref{eq:reid} and \eqref{CRB}, YFG have succeeded in establishing a link between the EPR paradox and the metrological power of a state~\cite{YBG}. They consider Bob's conditioned states $\rho^B_{k\vert K}$ to compute the optimised conditional variance
\begin{equation}
    \Delta^2 H_{\rm opt} = \min_K \sum_k p(k\vert K) \Delta^2 H_{k|K},
\end{equation}
and the quantum conditional Fisher information
\begin{equation}
\label{eq:qfisher}
    F_{\rm opt} = \max_K \sum_k p(k\vert K) F_{Q,k|K},
\end{equation}
where $B_{k|K}$ is the value of the quantity $B$ calculated on the state $\rho^B_{k\vert K}$. For any state admitting a decomposition in assemblages $\mathcal{A}(k,K)$, there holds the limit
\begin{equation}
\label{eq:YFG}
    F_{\rm opt} \leq 4\, {\Delta^2 H_{\rm opt}}.
\end{equation}
A violation of such an inequality flags the inappropriateness of local hidden state models, hence the presence of an EPR paradox. This can be put in direct connection to \eqref{eq:reid} and \eqref{CRB} by means of the Cram\'er-Rao bound.
The bounds on the optimal variances $\Delta^2H_{\rm opt}\leq \Delta^2H_{\rm est}$ are loose, and can not be employed to assess violation of nonsteering conditions.  We can proceed by inspecting directly the conditional variances
\begin{equation}
    \Delta^2H_{\rm cond} = \sum_{k} p(k\vert K)\Delta^2 H_{ k\vert K},
\end{equation}
with specific choices of the observables. By definition, $\Delta^2 H_{\rm opt}\leq \Delta^2 H_{\rm cond}$, ensuring the correct chain of inequalities.
and of the property that ${\Delta^2H_{\rm opt}}\leq{\Delta^2H_{\rm est}}$.

We tested these ideas in the experiment illustrated in Fig.~\ref{fig:apparato}. Bipartite states with a variable degree of steering are produced as follows. Two photons are prepared in the states $\vert \tilde D \rangle = \cos(\pi/3)\vert H \rangle + \sin(\pi/3)\vert V \rangle$, and $\vert\alpha\rangle=\cos(2\alpha)\vert H\rangle+\sin(2\alpha)\vert V \rangle$ - the kets $\vert H \rangle$ and $\vert V \rangle$ denote a photon with horizontal and vertical polarisation respectively. These arrive on a partially polarising beam splitter (PPBS), with transmittivities $T_H=1$, $T_V=1/3$, on which non-classical interference occurs. By postselecting events when the two photons emerge from different arms, a quantum correlated state is produced~\cite{PhysRevLett.95.210506}. The contributions in the final states are also modulated by the different transmittivities, and these have to be accounted for in the first preparation step: this implies that maximal correlation is expected for $2\alpha=\pi/3$. The two photons are then distributed to two measurement stations, Alice and Bob, for the analysis: a conventional sequence of a quarter wave plate and a half wave plate followed by a polariser implement the measurement of one of the Pauli operators $X,Y$ or $Z$. These correspond to discrimination of diagonal ($D$)/antidiagonal ($A$) polarisations for $X$, right-circular ($R$)/left circular ($L$) for $Y$, and $H$/$V$ for $Z$.

\begin{figure}
\centering
\begin{subfigure}[b]{\columnwidth}a)
\includegraphics[width=\columnwidth]{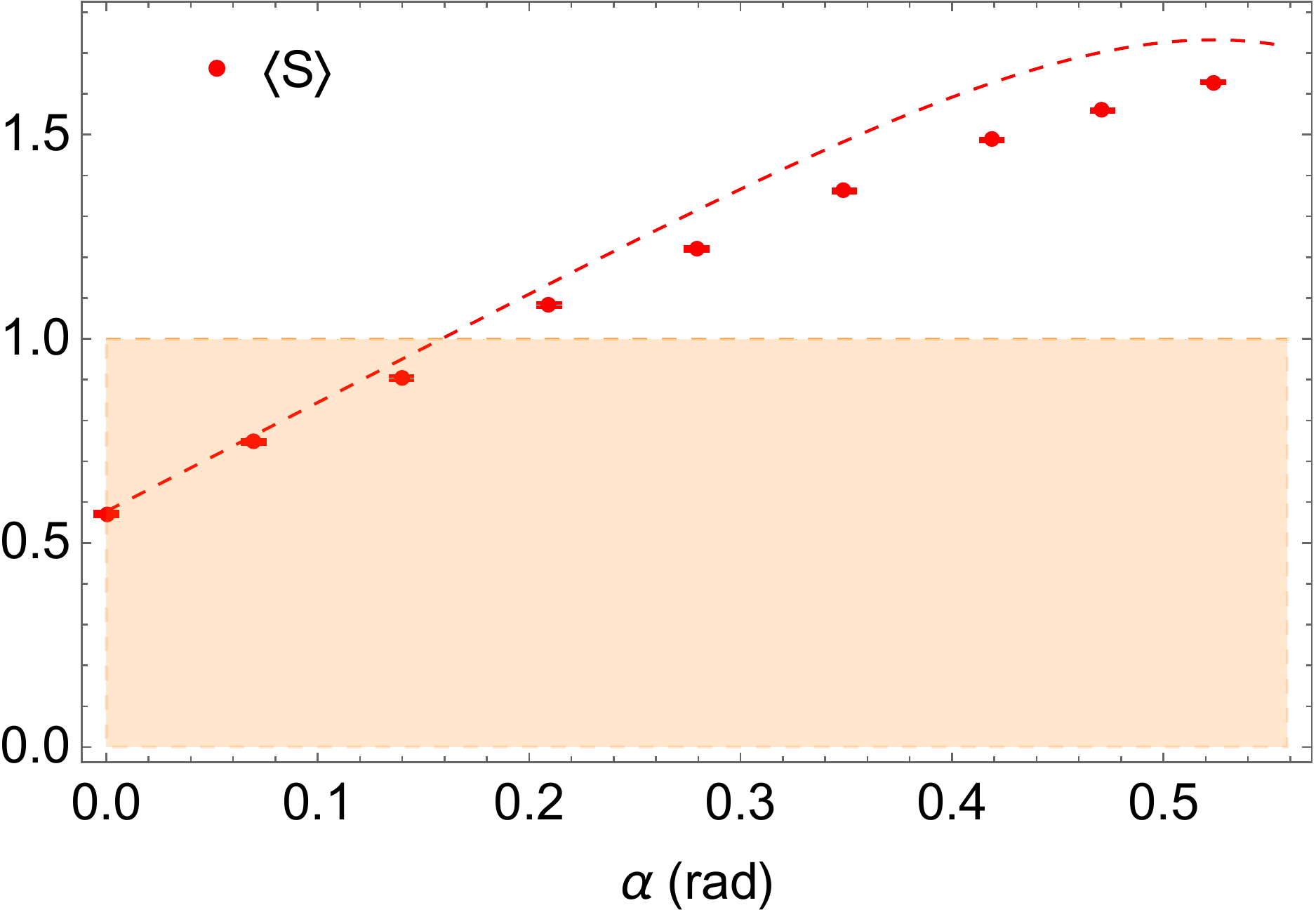}
    \end{subfigure}
    
\vskip 3.5 mm
    
\begin{subfigure}[b]{\columnwidth}b)
\includegraphics[width=\columnwidth]{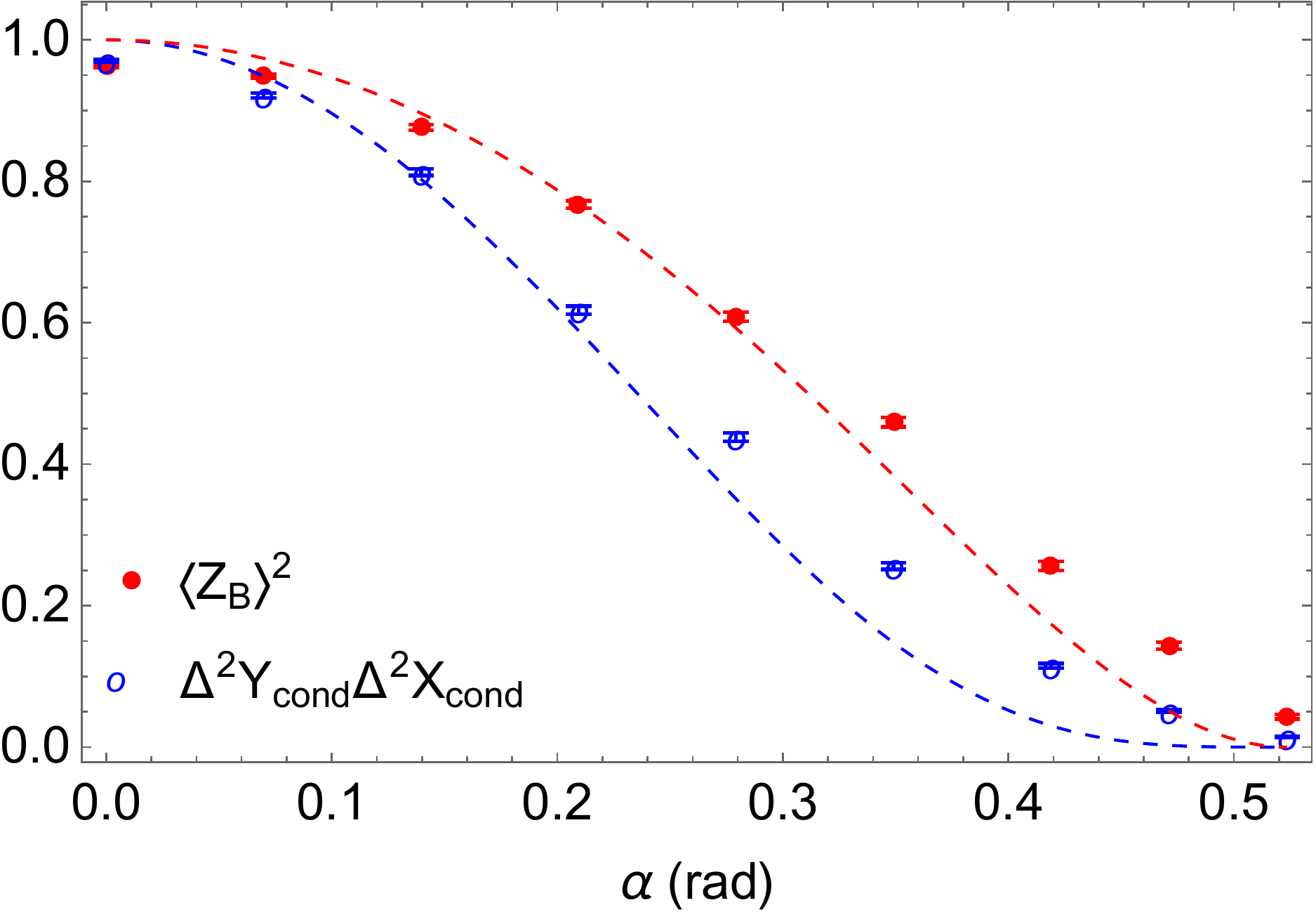}
    \end{subfigure}
 \caption{{\it Steering witness.} a. Assessment of the presence of steering by means of Eq. \eqref{eq:esse}, showing $\langle S \rangle$ as a function of the angle $\alpha$. The experimental results (red dots) are shown together with the ideal case (dashed red). The orange area indicates values of $\langle S \rangle\leq 1$.
 b) Test of the steering using Reid's criterion \eqref{eq:reid} applied to the conditional variances. The product of the conditional variances (blue) is well below the non-steerable measured bound (red), for every $\alpha$. 
 The points correspond to the measured values, while the dashed lines are predictions for the ideal state. 
 In both panels, the errors are evaluated by propagation of the Poisson statistics of the registered counts.}

    \label{fig:sparameter}
\end{figure}

We detect the presence of steering in Alice and Bob's shared state. The first method we adopt based on~\cite{PhysRevLett.98.140402,PhysRevA.80.032112}, considers a bound on correlations established for non-steerable state. 
For our class of states, this takes the form
\begin{equation}
\label{eq:esse}
    \langle S \rangle =\frac{1}{\sqrt{3}}\vert \langle X_A Z_B \rangle+\langle Z_A X_B \rangle+\langle Y_A Y_B \rangle \vert \leq 1. 
\end{equation}
Alice's measurement of $Z_A$ thus prepares Bob's photon in the $X_B$ basis, and similarly for the other two cases. While we have adopted a fully quantum notation for a direct connection to the measured quantities, in Bob's point of view, the expectation values are calculated as average correlations between Alice's classical input and his own quantum observable. The corresponding results are pictured in Fig.~\ref{fig:sparameter}, where $\langle S \rangle$ is reported as a function of the angle $\alpha$. For low values of $\alpha$ the existing correlations do not allow to confidently assess the presence of steering, while for higher values, this is present, despite the non-idealities of experiment, foremost the reduced two-photon interference visibility.

Consider the observables $H=Y_B$ and $H'=X_B$: Alice should implement a measurement of $K=Y_A$ and $k+Z_A$, respectively, in order to obtain the best predictions of Bob's outcomes. By this choice, the expressions of the variances in~\eqref{eq:deltaest} for $Y_B$ and $X_B$ can be recast as $\Delta^2Y_{\rm est}=1-\langle Y_A Y_B \rangle$, and $\Delta^2X_{\rm est}=1-\langle Z_A X_B \rangle$. According to Reid's criterion \eqref{eq:reid}, their product is bounded in non-steerable states by $\langle Z_B \rangle^2$, obtained by tracing out Alice's photon. The corresponding experimental results are reported in Fig.~\ref{fig:sparameter}b, and demonstrate a violation of the inequality for almost all states -- exception are the first point, for which this is expected, and the second, due to imperfections. We remark, however, that the amount of violation is not proportional to the level of steering, as it appears evident from the comparison of the two panels in Fig.~\ref{fig:sparameter}.
We consider these results as the set standard for a comparison with the metrological study.

\begin{figure}[h!]
\begin{subfigure}[b]{\columnwidth}a)
\includegraphics[width=\columnwidth]{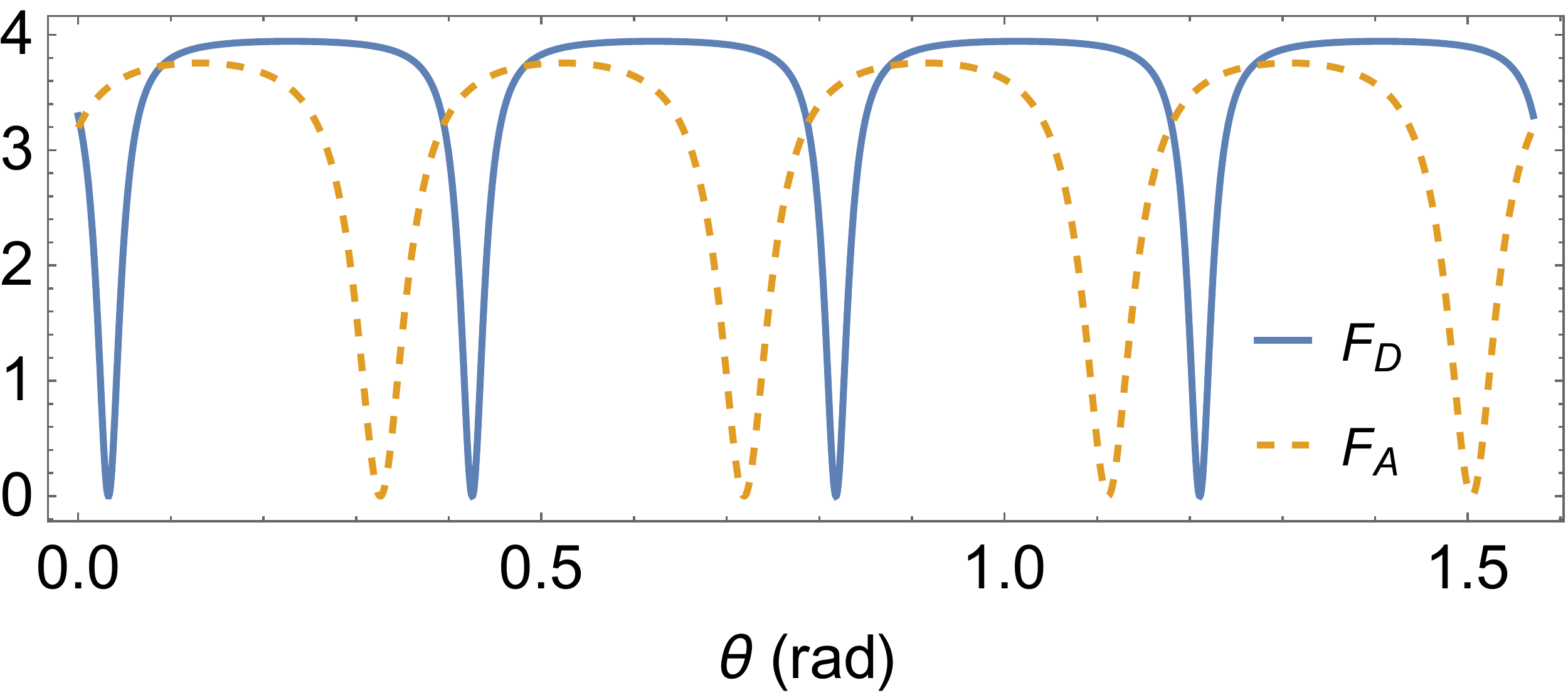}
\end{subfigure}

\vskip 3 mm

\begin{subfigure}[b]{\columnwidth}b)
\includegraphics[width=\columnwidth]{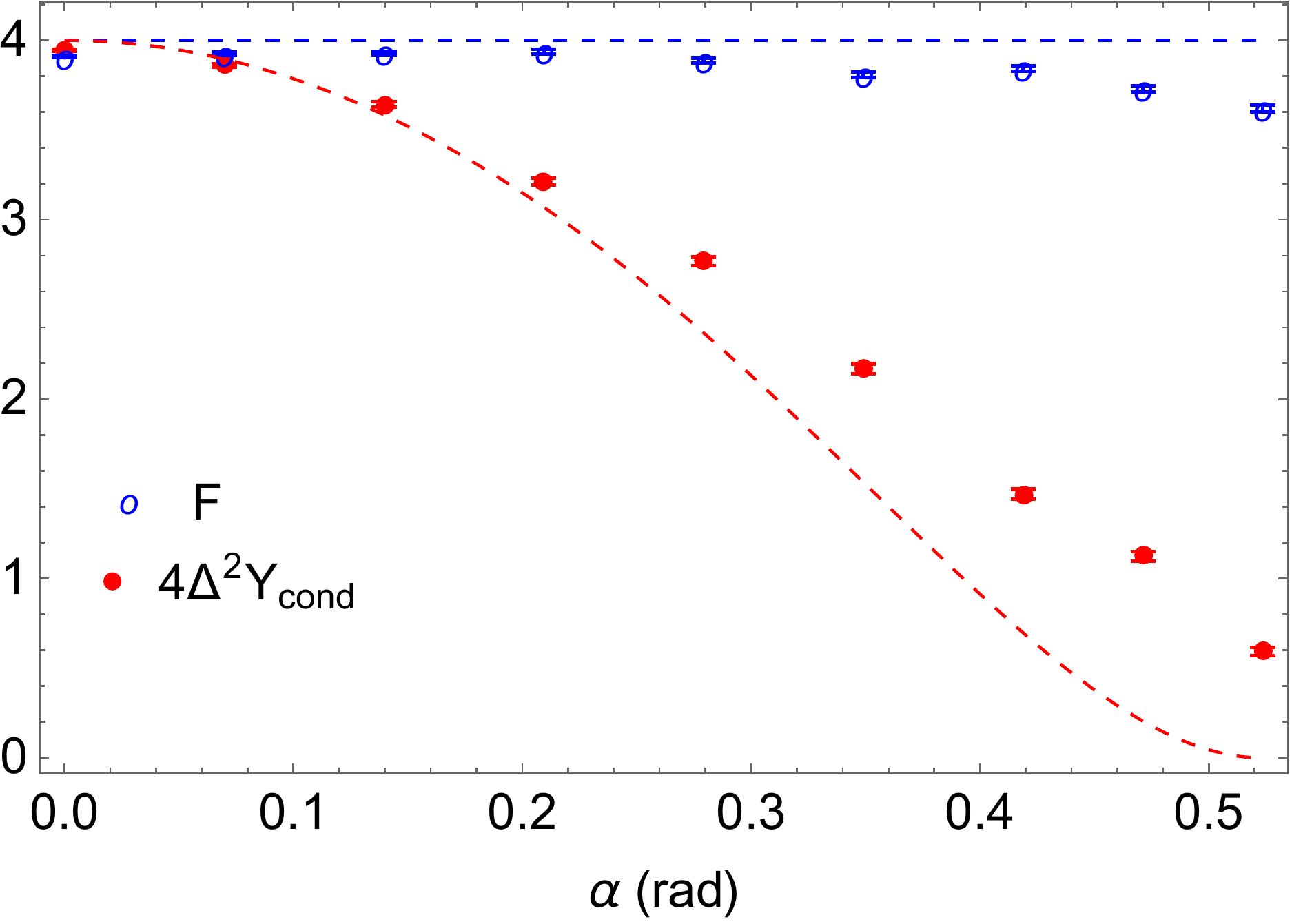}
\end{subfigure}

    \caption{{\it Fisher information.} a) Conditional Fisher information  ($F_D$ solid blue, $F_A$ dashed yellow) as a function of $\theta$ for $\alpha=0.42$ rad. b) measured variance (red) and quantum Fisher information (blue) as a function of $\alpha$:  the points correspond to the measured values, while the dashed lines are predictions for the ideal state. The errors are evaluated through a Monte Carlo routine with 100 runs.}
    \label{fig:duefisher}
\end{figure}

If Bob is convinced of the absence of steering in the distributed state, he would conclude the variance $\Delta^2Y_{\rm est}$ sets the achievable precision in an experiment aiming at estimating the parameter $\theta$ with the evolution $e^{i\theta Y_B}$. For instance, upon observing $\Delta^2Y_{\rm est}=0$, Bob would infer that the local hidden state collection he is receiving is equivalent to an incoherent mixture of eigenstates of $Y_B$, {\it i.e.} $\vert R \rangle$ and $\vert L \rangle$. However, Alice has the capability of steering Bob's state to $\vert H \rangle$ and $\vert V \rangle$: contrary to his expectation, Bob could benefit from these coherent superpositions of $\vert R \rangle$ and $\vert L \rangle$ for phase estimation.  

In our experiment, we have implemented the operation $e^{i\theta Y_B}$ by means of a half wave plate, set at an angle $\theta/2$, and recorded detection probabilities $p(H|D), p(V|D)$ and $p(H|A), p(V|A)$ in the $Z_B$ basis for Bob, conditioned by a measurement of $X_A$, {\it i.e.} of the $D$ or $A$ polarisations on Alice's side. The phase $\theta$ has been varied from 0 to $48^{\circ}$ in steps of $4^{\circ}$.The corresponding probabilities are then fitted by an expression of the kind $p(\theta) = (1+v\cos(\theta+\theta_0))/2$, with $v$ and $\theta_0$ being the fitting parameters. This allows to calculate the conditional Fisher information $F_D$ by means of its definition $F_D = \left(\partial_\theta p(H|D) \right)^2/p(V|D)+\left(\partial_\theta p(V|D) \right)^2/p(V|D)$, and similarly for $F_A$; typical results are reported in Fig.~\ref{fig:duefisher}a. This procedure is compatible with the assumption made upfront that Bob does not question the validity of quantum mechanics for the description of his apparatus - this is implicit in the choice of the fitting function.

The maximal values of $F_D$ and $F_A$ are used to compute a lower bound for the quantum Fisher information \eqref{eq:qfisher}; if the state is not steerable, this is expected to have $4\Delta^2Y_{\rm est}$ as upper limit. The violation of this condition is reported in Fig.~\ref{fig:duefisher}b: the solid red points indicate the measured variance, while the open blue points indicate the Fisher information and sit well above the non-steerable limit in the whole range of $\alpha$.

\section{Discussion}

Our results indicate the YFG and Reid's criterion offer consistent information on the presence of steering. Differently from the latter, YFG also provides a quantitative indication on the level of steering in the class of states we have tested; nevertheless it is unclear whether this properties holds in the general case~\cite{YBG}. This information is also captured by the steering witness $\langle S \rangle$, but this is attained only when exceeding a minimum level of steering. Pursuing the YFG route is demanding in terms of resources: obtaining an estimation of the Fisher information requires performing several measurements, and this compares unfavourably to the more economical method of Fig.~\ref{fig:sparameter} a). On the other hand, the closer inspection of Bob's state reveals an EPR paradox for a wider range of values of $\alpha$. The method preserves its conceptual importance, but appears less appealing for technological purposes, such as verification of steering in networks, especially when aiming at loophole-free arrangements. On the other hand, for many-particle atomic systems, a measurement of the Fisher information represent a convenient alternative~\cite{Strobel424}, and the method may find concrete applications. 

\section*{Acknowledgments} 
This   work   was   supported   by   the European  Commission  through  the  FET-OPEN-RIA  project STORMYTUNE (G.A. number 899587).

\bibliography{steeringFisher.bib}

\end{document}